\documentclass{article}

\usepackage[ruled,longend,linesnumbered]{algorithm2e}
\SetAlFnt{\small}
\SetAlCapFnt{\small}
\SetAlCapNameFnt{\small}
\SetAlCapHSkip{0pt}
\IncMargin{-\parindent}

\usepackage{graphicx}
\usepackage[caption = false]{subfig}

\usepackage{url}

\def\be{\begin{equation}}
\def\ee{\end{equation}}
\def\ba{\begin{array}}
\def\ea{\end{array}}
\def\bea{\begin{eqnarray}}
\def\eea{\end{eqnarray}}
\def\beas{\begin{eqnarray*}}
\def\eeas{\end{eqnarray*}}

\newcommand{\bfb}{\mbox{\boldmath $b$}}
\newcommand{\bfc}{\mbox{\boldmath $c$}}
\newcommand{\bfd}{\mbox{\boldmath $d$}}

\newcommand{\bfu}{\mbox{\boldmath $u$}}

\newcommand{\bfw}{\mbox{\boldmath $w$}}
\newcommand{\bfx}{\mbox{\boldmath $x$}}

\newcommand{\bfA}{\mbox{\boldmath $A$}}

\newcommand{\bfJ}{\mbox{\boldmath $J$}}

\newcommand{\bfU}{\mbox{\boldmath $U$}}
\newcommand{\bfV}{\mbox{\boldmath $V$}}

\newcommand{\bfxi}{\mbox{\boldmath $\xi$}}

\newcommand{\bfphi}{\mbox{\boldmath $\phi$}}

\title{Finite element numerical integration \\ for first order approximations \\ on multi-core architectures}
\author{Krzysztof Bana\'{s}\footnotemark[1]\ \footnotemark[3]
\and Filip Kru\.{z}el\footnotemark[2]\ \footnotemark[3]
\and Jan Biela\'{n}ski\footnotemark[1]\ \footnotemark[3]}

\begin{document}
	
\maketitle

\renewcommand{\thefootnote}{\fnsymbol{footnote}}

\footnotetext[1]{Department of Applied Computer Science and Modelling, 
AGH University of Science and Technology, Mickiewicza 30, 
30-059 Krak\'{o}w, Poland}
\footnotetext[2]{Institute of Computer Science, 
Cracow University of Technology, Warszawska 24, 31-155 Krak\'{o}w, Poland}
\footnotetext[3]{ The work presented in the paper was supported by the Polish National Science Centre under grant no DEC-2014/13/B/ST8/03812.}

\renewcommand{\thefootnote}{\arabic{footnote}}

\begin{abstract}
The paper presents investigations on the implementation and performance of the finite element numerical integration algorithm for first order approximations and three processor architectures, popular in scientific computing, classical CPU, Intel Xeon Phi and NVIDIA Kepler GPU. A unifying programming model and portable OpenCL implementation is considered for all architectures. Variations of the algorithm due to different problems solved and different element types are investigated and several optimizations aimed at proper optimization and mapping of the algorithm to computer architectures are demonstrated. Performance models of execution are developed for different processors and tested in practical experiments. The results show the varying levels of performance for different architectures, but indicate that the algorithm can be effectively ported to all of them. The general conclusion is that the finite element numerical integration can achieve sufficient performance on different multi- and many-core architectures and should not become a performance bottleneck for finite element simulation codes.  
\end{abstract}


%


\section{Introduction}

Finite element numerical integration forms an indispensable part of practically all complex finite element codes. Apart from special formulations for specific problems (such as e.g. applications in structural mechanics for beams, plates etc.) where it can be avoided using some analytically obtained formulae, it is universally used for creating entries to finite element stiffness matrices and load vectors (matrices and right hand side vectors for systems of linear equations). 

The recent progress in hardware for numerical computations caused the widespread use of multi-core and massively multi-core (often termed many-core) processors with vector processing capabilities. These created the necessity to reconsider the implementations of scientific codes, taking into account the possible options for multi-threading and the changing execution environment characteristics. 

The aim of the research presented in the current paper is to thoroughly analyse the process of finite element numerical integration for first order approximations, in a possibly broad context, and to propose a set of guidelines for designing efficient integration procedures for multi-core processors, taking into account SIMD modes of execution. The paper shows also some variants of the algorithm, that are implemented for different hardware platforms and tested in practice.   

The paper is organised as follows. First, finite element numerical integration is described in the form investigated in our work. Then, in Section 3 its implementation, in the context of finite element simulations and for different types of problems, elements and computing platforms is analysed. Several variants of the procedure are developed, and, in the next section, their  optimization and mapping to different processor architectures is discussed. Section 5 presents the results of computational experiments that test the performance of procedures with short discussion of results. The last two sections briefly document other works related to the subject presented in the current paper and present the final conclusions.

\section{Finite element numerical integration}

\subsection{Finite element weak statements and systems of linear equations}
In the current paper, we consider the finite element method as a tool for finding approximate solutions to partial differential equations, specified in a computational domain $\Omega$ with suitable boundary conditions imposed on $\partial \Omega$, based on weak formulations that can be expressed (with some details omitted) as follows \cite{johnson_book}:
\vspace*{2mm}

Find an unknown function $\bfu$ belonging to a specified function space $\bfV^h$, such that 
 \bea 
\int_{\Omega} \left(
\bfc^{i;j}  \bfw_{,i} \bfu_{,j}
+
 \bfc^{i;0} \bfw_{,i}  \bfu
+
 \bfc^{0;i} \bfw  \bfu_{,i}
+
   \bfc^{0;0} \bfw  \bfu \right) 
d\Omega + \rm{BTL}
 =
\label{weak_form} \\ 
= \int_{\Omega} \left(
 \bfd^{0} \bfw 
 +  
 \bfd^{i} \bfw_{,i}  \right) d\Omega
+ \rm{BTR} 
\nonumber
\eea
for every test function $\bfw$  defined in  a space $\bfV^h_0$.

\vspace*{2mm}
Above, $\bfc^{i;j}$ and $\bfd^{i}$, $i,j=0,..,N_D$ denote the coefficients specific to the weak formulation (with $N_D$ being the number of physical space dimensions), ''$,$'' denotes differentiation with respect to space coordinates and summation convention for repeated indices is used. $\rm{BTL}$ and $\rm{BTR}$ denote boundary terms, usually associated with integrals over the boundary $\partial \Omega$. The task of calculating boundary terms is usually much less computationally demanding, and we neglect it in the current paper, when discussing implementations for multi- and many-core architectures. In the actual computations described later in the paper this task is performed by the part of the code executed by CPU cores, using standard finite element techniques.


There are many possible finite element approximation spaces that can be defined for triangulations of $\Omega$ into a set of finite elements $\Omega_{e}$, $e=1,...,N_E$ 
In the current paper we concentrate on scalar problems and the popular case of first order approximations, i.e. spaces spanned by elementwise linear (or multilinear) basis functions, associated with element vertices. Vertices with the associated basis functions form the nodes of the finite element mesh, with their number denoted by $N_N$. Piecewise linear basis functions are defined for simplex elements (triangles in 2D and tetrahedra in 3D), multilinear basis functions are specified for other types of elements (quadrilateral, prismatic, hexahedral, etc.) 


After representing unknown and test functions as linear combinations of basis functions $\psi^r$, $r=1,..,N_N$,
the system (\ref{weak_form}) is transformed to a set of linear equations for the unknown vector of degrees of freedom $\bfU$:
\be
\sum_{s=1}^{N_N} A^{rs} \cdot U^s = b^r \hspace{15mm} r=1,..,N_N
\label{linear_system_global}
\ee 
with the entries in the global system matrix $\bfA$ (the stiffness matrix) computed as (with boundary terms neglected):
\be
{A}^{rs} \! = \! \int_{{\Omega}} \left(
\sum_{i} \sum_{j} c^{i;j} \psi^r_{,i} \psi^s_{,j}
+
\sum_{i} c^{i;0} \psi^r_{,i} \psi^s_
+
\sum_{i} c^{0;i}  \psi^r \psi^s_{,i}
+
   c^{0;0}  \psi^r \psi^s \right) 
d\Omega
\label{sm_global}
\ee

The entries to the global right hand side vector $\bfb$ (the load vector) are calculated according to the formula (again with boundary terms neglected):
\be
{b}^{r} \! = \! \int_{{\Omega}} \left(
\sum_{i} d^{i}  \psi^r_{,i}
+
d^{0}  \psi^r  \right) 
d\Omega
\label{lv_global}
\ee 

The solution of a single problem (\ref{weak_form}) (or equivalently the system (\ref{linear_system_global})) may constitute the only step of the finite element solution process or may form a single stage of more complex solution strategies, e.g. for time dependent and/or nonlinear problems. In the latter case the efficient accomplishment of numerical integration becomes even more important for the overall performance of simulation programs.



\subsection{Finite element solution procedures}
The solution of a single finite element problem represented by  (\ref{weak_form}) or  (\ref{linear_system_global}) requires the creation of global stiffness matrix and load vector entries, followed by their use in obtaining the final solution of the problem.
A single entry in the global stiffness matrix or load vector consists of an integral over the whole computational domain plus some boundary terms. The integral over the computational domain, on which we concentrate in the current paper, can be expressed as the sum of contributions from individual elements:
\be
 \int_{{\Omega}} ... d\Omega = \sum_{e}  \int_{{\Omega}_{e}} ... d\Omega
\ee

The entries can be either computed individually, in a double loop over all entries of the global stiffness matrix, or can be computed in a loop over all elements, with all the entries related to a given element computed at once. 
We consider in the current paper only the latter strategy.
Compared to the first strategy where each finite element is visited several times, this saves some element related calculations but leaves as a result element contributions that have to be further processed. Our choice is motivated by the fact that for many types of problems, including nonlinear, the overhead of repeated element calculations may become too high. In our approach we do not decide, however, how element contributions are used further in the solution procedure. Different options, with the assembly of the global stiffness matrix \cite{Cecka_2011} or without assembly, in the so called matrix-free approaches \cite{matrixfree_arbenz}, can be applied.

The element contributions to the global stiffness matrix form a small dense matrix ${\bfA}^{e}$, with each entry corresponding to a pair of element shape functions $\phi$ (that are used for creation of specific global basis functions).
The entries of the element stiffness matrix (with boundary terms neglected) are given by the formula:
\be
({A}^{e})^{rs} \! = \! \! \int_{{\Omega}_{e}} \! \! \left(
\sum_{i} \sum_{j} c^{i;j}  \phi^r_{,i} \phi^s_{,j}
\! + \!
\sum_{i} c^{i;0}  \phi^r_{,i} \phi^s
\! + \!
\sum_{i} c^{0;i}   \phi^r \phi^s_{,i}
\! + \!
 c^{0;0}  \phi^r \phi^s \! \right) 
\! d\Omega
\label{sm_local}
\ee 
where now indices $r$ and $s$ are local indices, with the range from 1 to $N_S$ -- the number of shape functions for a given element.

The element right hand side vector, that forms the element contribution to the global right hand side vector, is computed based on the formula (with boundary terms neglected):
\be
({b}^{e})^{r} = \int_{{\Omega}_{e}} \left(
\sum_{i} d^{i} \phi^r_{,i}
+
   d^{0}  \phi^r  \right) 
d\Omega 
\label{lv_local}
\ee 
 

\subsection{A generic finite element numerical integration algorithm}
We consider in our paper the creation of element stiffness matrices and load vectors using numerical integration and transformations to reference elements. Both procedures are related, since quadratures for numerical integration are usually defined for reference elements.

Shape functions are defined for a reference element of a given type of geometry and approximation. Each real element in the finite element mesh is treated as an image of the associated reference element under a suitable geometric transformation. Denoting the physical coordinates of points in the real element by $\bfx$, the transformation (defined separately for each element)  from the reference space with coordinates $\bfxi$ is expressed as $\bfx(\bfxi)$ (with the inverse transformation given by $\bfxi(\bfx)$). 

The parameters of the transformation are used for calculations of
global derivatives of shape functions that involve the elements of the Jacobian matrix of the inverse transformation $\bfxi(\bfx)$ (the elements of Jacobian matrices 
$\left\{\frac{\partial{\bfx}}{\partial{\bfxi}}\right\}$ and
$\left\{\frac{\partial{\bfxi}}{\partial{\bfx}}\right\}$ with their determinants, are later on, when considering actual computations, termed together as Jacobian terms).

The transformation from the reference element $\hat{\Omega}$ to the real element ${\Omega}_{e}$ is used to perform the change of variables in integrals from  (\ref{sm_local}) and (\ref{lv_local}). Taking an example integral from (\ref{sm_local}) and applying the change of variables we obtain:
\be
\! \! \int_{{\Omega}_{e}} \! \!
\sum_{i} \! \sum_{j}  \! c^{i;j}  \frac{\partial  \phi^r}{\partial x_i} 
 \frac{\partial \phi^s}{\partial x_j} d\Omega
= 
  \int_{\hat{\Omega}} 
 \sum_{i} \! \sum_{j}  \! c^{i;j} 
  \sum_k \! \frac{\partial \hat \phi^r}{\partial \xi_k} \frac{\partial \xi_k} {\partial x_i} 
 \sum_z \! \frac{\partial \hat \phi^s} {\partial \xi_z} \frac{\partial \xi_z} {\partial x_j}  
 \det \bfJ d\Omega
\label{change_of_variables}
\ee
where $\hat {\phi^r}$ denotes the shape functions defined for the reference element. 

The application of numerical integration quadratures (in practical examples we always use Gaussian quadratures) leads to the transformation (for brevity we use global derivatives of shape functions):
\bea
\!\! \! \! \! \int_{{\Omega}_{e}} \!
\sum_{i} \sum_{j} c^{i;j}  \frac{\partial  \phi^r}{\partial x_i} 
 \frac{\partial \phi^s}{\partial x_j} d\Omega
& \approx &
\sum_{Q=1}^{N_Q} \!
\left. \left(
\sum_{i} \sum_{j}  c^{i;j}  
 \frac{\partial  \phi^r} {\partial x_i} 
  \frac{\partial  \phi^s}  {\partial x_j}  
 \det \bfJ 
\right) \right|_{\bfxi^Q}
\! w^Q
 \label{quadrature}
\eea
where $\bfxi^Q$ denotes the coordinates of the subsequent integration points with the associated weights $w^Q$, $Q=1,..,N_Q$ ($N_Q$ being the number of integration points related to the type of element, the degree of approximation and the form of coefficients $c^{i;j}$). 



The input for the generic procedure of numerical integration for a single real element, that implements (\ref{sm_local}) and (\ref{lv_local}), with (\ref{change_of_variables}), (\ref{quadrature}) and similar expressions for other terms taken into account, is given by:
\begin{itemize}
\item
the type of element, the type and the order of approximation
\item
the set of local coordinates and weights for integration points
\item
possibly the set of values of shape functions and their local derivatives at all integration points for the reference element (they can be also computed on the fly during integration)
\item
the set of geometry degrees of freedom for the real element
\item
the set of problem dependent parameters that are used for computing coefficients $c^{i;j}$ and $d^{i}$ at integration points (for simple problems these may be the values of coefficients at integration points themselves)
\end{itemize}

\label{specification_of_numerical_integration}

The output of the procedure is formed by the element stiffness matrix ${\bfA}^{e}$ and the element load vector ${\bfb}^{e}$. Algorithm \ref{num_int_generic} shows a generic procedure for creating a sequence of element stiffness matrices and load vectors, assuming they have the same type and order of approximation, with the data for the respective reference element (integration data and shape functions data) assumed to be available in local data structures.

\begin{algorithm}
\For{$e=1$ to $N_E$}{
- read problem dependent parameters specific to the element 
 \\
- read geometry data for the element \\
- initialize ${\bfA}^{e}$ and ${\bfb}^{e}$ 
\\
\For{$i_Q=1$ to $N_Q$} {
- compute necessary terms of Jacobian matrices $\frac{\partial \bfx}{\partial \bfxi}$ and $\frac{\partial \bfxi}{\partial \bfx}$, and calculate the value of variable {\bf vol}$[i_Q]$ = $\det(\frac{\partial \bfx}{\partial \bfxi})\times  \bfw^Q [i_Q]$ \\
\For{$i_{S}=1$ to $N_{S}$}{
- using Jacobian terms calculate the values of global derivatives of shape functions at integration points,  $\bfphi$ \\
}
}
\For{$i_Q=1$ to $N_Q$} {
- calculate the actual coefficients  $\bfc$ and  $\bfd$ 
at integration points \\
}
\For{$i_Q=1$ to $N_Q$} {
\For{$i_{S}=1$ to $N_{S}$}{
\For{$j_{S}=1$ to $N_{S}$}{
\For{$i_D=0$ to $N_D$}{
\For{$j_D=0$ to $N_D$}{
${\bfA}^{e}[i_{S}][j_{S}]+= $ \\ $\; \; \; \; \; \; \; \; $
{\bf vol}$[i_Q]$ $\times \bfc[i_D][j_D][i_Q] \times \bfphi[i_D][i_{S}][i_Q] \times \bfphi[j_D][j_{S}][i_Q]$ \\
} 
} 
} 
\For{$i_D=0$ to $N_D$}{
${\bfb}^{e}[i_{S}]+= $
{\bf vol}$[i_Q]$ $\times \bfd[i_D][i_Q] \times \bfphi[i_D][i_{S}][i_Q] $ \\
} 
} 
} 
- store the whole ${\bfA}^{e}$ and ${\bfb}^{e}$ 
as output of the procedure
} 
\caption{A generic algorithm for finite element numerical integration for the sequence of elements of the same type and order of approximation}
\label{num_int_generic}
\end{algorithm}

The algorithm shows how numerical integration calculations can be performed. 
As it can be seen, there are three subsequent, separate loops over integration points in the algorithm. They can be fused together, as it is often done in practice. The presented form of Algorithm  \ref{num_int_generic} has been chosen to show that, with the suitable arrangement of calculation stages, the final computations of stiffness matrix entries are performed in five nested loops, with the order of loops being arbitrary. Each order can, however, lead to different execution performance, that additionally depends on the hardware on which the code runs. This opens multiple ways for optimizing the execution characteristics of finite element numerical integration that we investigate in the following sections.

\section{Implementation of numerical integration}

\subsection{The place of numerical integration within the finite element solution procedure}
We treat the procedure of numerical integration as one of computational kernels of finite element codes. By this we want to stress its importance for finite element computations and its influence on the performance of codes. To further clarify its place in finite element programs we describe a way in which, in the setting assumed in the paper, it interacts with the rest of the code.

We consider the implementation of finite element numerical integration in codes designed for large scale calculations. The number of elements in such calculations often exceeds million and even can reach several billions. The only way to ensure the scalable execution of programs for that size of problems is to consider distributed memory systems and message passing. The standard way of implementation in this case is some form of domain decomposition approach. In the current paper we assume the design of finite element codes, in which the issues of message passing parallelization and multi-core acceleration are separated \cite{iccs04}. In this design multithreading concerns only single node execution and a separate layer of software combines single node components into a whole parallel code.

As a result we consider the algorithm of numerical integration executed for all elements of a given subdomain. We assume the size of the subdomain to exceed one hundred thousand of elements, the size that justifies the use of massively multi-core accelerators. 
We allow for complex mesh and field data structures related to adaptive capabilities, many types of elements, several approximation fields and different approximation spaces in a single simulation. Therefore we assume that the finite element data concerning elements and approximation fields are stored in standard global memory accessible to CPUs. This memory should ensure the sufficient storage capacity and a latency minimizing way of accessing data.

Hence, the input data to numerical integration comes from finite element data structures stored in CPU memory. The output of numerical integration is formed by element stiffness matrices and load vectors. In order to allow for broad range of scenarios of using the computed entries (standard direct, frontal, iterative solvers, with or without matrix assembly) we assume that the output arrays after being computed, can be stored in some memory, available to threads performing subsequent solution steps (this can even include registers if e.g. assembly operation is designed as extension of procedures performing numerical integration). Hence, we discuss the problem of effective storing of element arrays in memory of the processor performing integration, but consider also the case, where the data are not stored at all. In both cases the further use of element stiffness matrices and load vectors depends on implementation details of possible assembly and solver procedures.



\subsection{Parallelization strategy}
As can be seen in Algorithm  \ref{num_int_generic}, numerical integration in the form considered in the current paper, consists of several loops. Each of these loops can be parallelized, leading to different parallel algorithms with different optimization options and performance characteristics.

For the analysis in the current paper we consider parallelization of only one loop -- the loop over finite elements. Due to the sufficiently large number of elements in the loop (as assumed for message passing implementation of large scale problems) and the embarrassingly parallel character of the loop this choice seem to be obvious. By analysing it in details, we want to investigate its optimization and performance characteristics, but also to show possible limitations that can lead to different parallelization strategies. 

Numerical integration is an embarrassingly parallel algorithm, but the later stages, assembly and linear system solution, are not. To allow for efficient execution of these stages, 
 further assumptions are made. As was described when developing Algorithm  \ref{num_int_generic}, elements of a single subdomain are divided into sets of elements of the same type and degree of approximation (to made them use the same reference element shape functions and integration quadratures). In order to allow for parallel assembly, the elements are further divided into subsets  with different colors assigned. Elements of a single color do not contribute to the same entries in the global stiffness matrix and their stiffness matrices can be easily assembled into the global stiffness matrix in a fully parallel way, with no data dependencies. 


\subsection{Variations of the algorithm}
Despite the fact that finite element numerical integration can be expressed by two simple formulas (\ref{sm_local}) and (\ref{lv_local}), there exist many variations of the algorithm (even when considering only first order approximations), that should be taken into account when designing computer implementations. 

First, there are different types of problems solved, scalar and vector, linear and non-linear. Depending on the weak statement of the corresponding problem, the set of PDE coefficients can be either sent as input to the procedure or the input to the procedure can form just the arguments of a special, problem dependent function, that calculates the final coefficients. In the current paper we consider the first option, with the assumption that the coefficients for element stiffness matrices are constant over the whole finite element. 

The coefficient matrices for different problems (matrices $\bf{c}$ in Algorithm \ref{num_int_generic}) can have varying non-zero structure. To make our implementations, to certain extent, independent of these variations, we consider the organization of computations with the final calculations of updates to the stiffness matrix (two loops over space dimensions  $i_D$ and $j_D$ -- lines 17-22 in Algorithm \ref{num_int_generic}) always performed inside all other loops. 
With such a solution, these final calculations are left as problem dependent and should be optimized separately, for each problem, either manually or using some automatic tools \cite{Luporini_2015}.

Another variation is related to different geometrical types of finite elements: linear, multilinear and geometrically higher order (e.g. isoparametric for higher order approximations).
The main difference, from the point of view of numerical integration, is between linear simplex elements, for which Jacobian terms are constant over the whole element, and all other types for which Jacobian terms have to be computed separately for each integration point. In our study we compare two types of elements: geometrically linear tetrahedral elements and geometrically multi-linear prismatic elements (both are elements without curved boundaries, with geometry degrees of freedom given by vertices coordinates). 

For the two different types of geometry of elements we consider different variations of numerical integration for first order approximations. We explicitly take into account the fact that for geometrically linear elements the derivatives of geometric shape functions and solution shape functions are constant over each element and that the corresponding calculations can be moved out of the loop over integration points.

%

Finally, we consider several options for arranging the order of calculations in Algorithm \ref{num_int_generic}. 
For the three loops, over integration points ($i_Q$) and twice over shape functions ($i_S$ and $j_S$), we consider the different orderings and denote them by QSS, SQS and SSQ (the symbol reflects the order implied by the subscripts of the loop indices). We assume that there are no separate loops for computing the additional quantities, Jacobian terms and global derivatives of shape functions, but that the loops are fused together. 

Combining together the options discussed above we present six algorithms (\ref{QSS_prism}, \ref{QSS_tetra}, \ref{SQS_prism}, \ref{SQS_tetra}, \ref{SSQ_prism}, \ref{SSQ_tetra})
for the final calculations of entries to  ${\bfA}^{e}$ (and ${\bfb}^{e}$ as well). The algorithms correspond to: 
\begin{itemize}
\item
different loop orderings -- QSS, SQS and SSQ versions
\item
different types of elements -- $geo\_linear$ for geometrically linear elements and $geo\_generic$ for all other types (it can be used for geometrically linear elements, but is less efficient)
\item
different types of problems -- through the delegation of final updates to problem specific calculations.
\end{itemize}

In the algorithms it is assumed that the values of Jacobian terms and the global derivatives of all shape functions are calculated just after entering the calculations for a given integration point.
The global derivatives are stored in some temporary array and later used in calculations. 
In practical calculations, in order to save the storage for temporary arrays, another strategy 
 can be used 
with the values of global derivatives for each shape function computed just at the beginning of  the body of the loop for this shape function. The drawback of the approach is that the calculations in the innermost loop over shape functions are redundantly repeated for each shape function.


\begin{algorithm}
\For{$i_Q=1$ to $N_Q$} {
calculate $\frac{\partial \bfxi}{\partial \bfx}$ and {\bf vol} (and possibly coefficients \bfc) \\
\For{$i_{S}=1$ to $N_{S}$}{
calculate  global derivatives of shape functions at the integration point \\
}
\For{$i_{S}=1$ to $N_{S}$}{
\For{$j_{S}=1$ to $N_{S}$}{
update
${\bfA}^{e}[i_{S}][j_{S}] $
using
$ \bfc[i_D][j_D][i_Q]$, $\bfphi[i_D][i_{S}][i_Q]$ and $\bfphi[j_D][j_{S}][i_Q]$ \\
} 
update ${\bfb}^{e}[i_{S}] $ using
$ \bfd[i_D][i_Q]$ and $\bfphi[i_D][i_{S}][i_Q]$ \\
} 
} 
store the whole ${\bfA}^{e}$ and ${\bfb}^{e}$ 
as output of the procedure
\caption{The $QSS\_geo\_generic$ version of numerical integration algorithm (explanation in the text)}
\label{QSS_prism}
\end{algorithm}

\begin{algorithm}
calculate $\frac{\partial \bfxi}{\partial \bfx}$ and {\bf vol} \\
\For{$i_{S}=1$ to $N_{S}$}{
calculate  global derivatives of shape functions  \\
}
\For{$i_Q=1$ to $N_Q$} {
(calculate possibly coefficients \bfc) \\
\For{$i_{S}=1$ to $N_{S}$}{
\For{$j_{S}=1$ to $N_{S}$}{
update
${\bfA}^{e}[i_{S}][j_{S}] $
using
$ \bfc[i_D][j_D][i_Q]$, $\bfphi[i_D][i_{S}][i_Q]$ and $\bfphi[j_D][j_{S}][i_Q]$ \\
} 
update ${\bfb}^{e}[i_{S}] $ using
$ \bfd[i_D][i_Q]$ and $temp[i_D] $ \\
} 
} 
store the whole ${\bfA}^{e}$ and ${\bfb}^{e}$ 
as output of the procedure
\caption{The $QSS\_geo\_linear$ version of numerical integration algorithm (explanation in the text)}
\label{QSS_tetra}
\end{algorithm}

\begin{algorithm}
\For{$i_{S}=1$ to $N_{S}$}{
initialize row of ${\bfA}^{e}$  \\
\For{$i_Q=1$ to $N_Q$} {
calculate $\frac{\partial \bfxi}{\partial \bfx}$ and {\bf vol} (and possibly coefficients \bfc)\\
calculate  global derivatives of shape functions at the integration point \\
\For{$j_{S}=1$ to $N_{S}$}{
update
${\bfA}^{e}[i_{S}][j_{S}] $
using
$ \bfc[i_D][j_D][i_Q]$, $\bfphi[i_D][i_{S}][i_Q]$ and $\bfphi[j_D][j_{S}][i_Q]$ \\
} 
update ${\bfb}^{e}[i_{S}] $ using
$ \bfd[i_D][i_Q]$ and $\bfphi[i_D][i_{S}][i_Q]$ \\
} 
store row of ${\bfA}^{e}$ and ${\bfb}^{e}[i_{S}] $ in global memory
} 
\caption{The $SQS\_geo\_generic$ version of numerical integration algorithm (explanation in the text)}
\label{SQS_prism}
\end{algorithm}

\begin{algorithm}
calculate $\frac{\partial \bfxi}{\partial \bfx}$ and {\bf vol} \\
\For{$i_Q=1$ to $N_Q$} {
calculate  global derivatives of shape functions \\
} 
\For{$i_{S}=1$ to $N_{S}$}{
initialize row of ${\bfA}^{e}$  \\
\For{$i_Q=1$ to $N_Q$} {
(calculate possibly coefficients \bfc) \\
\For{$j_{S}=1$ to $N_{S}$}{
update
${\bfA}^{e}[i_{S}][j_{S}] $
using
$ \bfc[i_D][j_D][i_Q]$, $\bfphi[i_D][i_{S}][i_Q]$ and $\bfphi[j_D][j_{S}][i_Q]$ \\
} 
update ${\bfb}^{e}[i_{S}] $ using
$ \bfd[i_D][i_Q]$ and $\bfphi[i_D][i_{S}][i_Q] $ \\
} 
store row of ${\bfA}^{e}$ and ${\bfb}^{e}[i_{S}] $ in global memory
} 
\caption{The $SQS\_geo\_linear$ version of numerical integration algorithm (explanation in the text)}
\label{SQS_tetra}
\end{algorithm}

\begin{algorithm}
\For{$i_{S}=1$ to $N_{S}$}{
\For{$j_{S}=1$ to $N_{S}$}{
${\bfA}^{e}[i_{S}][j_{S}] = 0$; ${\bfb}^{e}[i_{S}] = 0$ \\
\For{$i_Q=1$ to $N_Q$} {
calculate $\frac{\partial \bfxi}{\partial \bfx}$ and {\bf vol} (and possibly coefficients \bfc)\\
calculate  global derivatives of shape functions at the integration point \\
update
${\bfA}^{e}[i_{S}][j_{S}]$
using
$ \bfc[i_D][j_D][i_Q]$, $\bfphi[i_D][i_{S}][i_Q]$ and $\bfphi[j_D][j_{S}][i_Q]$ \\
\If{$i_{S}=j_{S}$}{
update ${\bfb}^{e}[i_{S}] $ using
$ \bfd[i_D][i_Q]$ and $\bfphi[i_D][i_{S}][i_Q] $ \\
} 
} 
store ${\bfA}^{e}[i_{S}][j_{S}] $ and ${\bfb}^{e}[i_{S}] $ (if $i_{S}=j_{S}$) in global memory
} 
} 
\caption{The $SSQ\_geo\_generic$ version of numerical integration algorithm (explanation in the text)}
\label{SSQ_prism}
\end{algorithm}

\begin{algorithm}
calculate $\frac{\partial \bfxi}{\partial \bfx}$ and {\bf vol} \\
\For{$i_{S}=1$ to $N_{S}$}{
calculate  global derivatives of shape functions \\
} 
\For{$i_{S}=1$ to $N_{S}$}{
\For{$j_{S}=1$ to $N_{S}$}{
${\bfA}^{e}[i_{S}][j_{S}] = 0$; ${\bfb}^{e}[i_{S}] = 0$ \\
\For{$i_Q=1$ to $N_Q$} {
(calculate possibly coefficients \bfc) \\
update
${\bfA}^{e}[i_{S}][j_{S}]$
using
$ \bfc[i_D][j_D][i_Q]$, $\bfphi[i_D][i_{S}][i_Q]$ and $\bfphi[j_D][j_{S}][i_Q]$ \\
\If{$i_{S}=j_{S}$}{
update ${\bfb}^{e}[i_{S}] $ using
$ \bfd[i_D][i_Q]$ and $\bfphi[i_D][i_{S}][i_Q] $ \\
} 
} 
store ${\bfA}^{e}[i_{S}][j_{S}] $ and ${\bfb}^{e}[i_{S}] $ (if $i_{S}=j_{S}$) in global memory
} 
} 
\caption{The $SSQ\_geo\_linear$ version of numerical integration algorithm (explanation in the text)}
\label{SSQ_tetra}
\end{algorithm}

The six variants of numerical integration can be considered as the optimizations of the algorithm, corresponding to classical techniques of loop interchange and loop invariant code motion. We perform them explicitly, to help compilers and to make the output for the same variant from different compilers similar. There are other different optimizations possible, such as common subexpression elimination, loop unrolling or induction variable simplification. In order to preserve the portability of the procedures for different processor architectures, we leave the application of these and other optimizations to the compiler. Instead, aiming particularly at optimization of the execution on graphics processors, we explicitly consider the placement of different arrays appearing in calculation in different levels of memory hierarchy. Before we present the details, we briefly introduce the programming model and example hardware platforms for which we design the implementations.

\subsection{Programming model}
We want to compare the optimizations and performance of execution for numerical integration on several types of hardware 
-- multi-core CPUs, GPUs, Xeon Phi -- 
but do not want to consider a radically different programming models. We choose a programming model that offers the explicit usage of shared memory (necessary for GPUs), but that can be used also for architectures based on standard CPU cores (we include in this category also Xeon Phi architecture). The model has to allow for efficient vectorization, as well as other standard optimizations.

The model that we choose is based on models developed for GPUs (CUDA \cite{CUDA_guide}, OpenCL \cite{OpenCL}), but tries to simplify them. The goal is to obtain the model simple enough, so that application programmers can use it, but having enough features to allow for high performance implementations on different contemporary hardware. Our final implementations are all done in OpenCL due to the fact of existing software development tools for all processors considered.

We assume that the code is written in the SPMD (single program multiple data) fashion, for threads that execute only scalar operations. Several threads can be grouped together, so that hardware can run them in lockstep, creating vectorised parts of the code. It is up to the compiler and the hardware to perform such vectorization. This vectorization is, in current programming environments, hidden from the programmer, but cannot be neglected when analysing the performance of implementation. The vectorization is considered the standard form of execution for GPUs and an important performance improvement for architectures with standard CPU cores. 

Because of the lockstep execution, we will call a group of threads performing vectorised operations a SIMD group (this term directly corresponds to the notions of NVIDIA \textit{warps} and AMD \textit{wavefronts}). The number of threads in a SIMD group depends on scheduling and execution details of each processor architecture. For the contemporary GPUs it is usually equal to 32 or 64, while for the current CPU like cores it is related to the width of vector registers and can be equal to 4, 8 or 16\footnote{In the model, we adopt the point of view on execution on standard CPU cores taken from Intel OpenCL compiler \cite{OpenCL_on_Xeon_Phi}. In that perspective, a traditional CPU thread performing vector operations is seen as a SIMD group of threads, while for scalar operations, it is assumed that they are performed by a single thread in the SIMD group.}.

One or several SIMD groups form a set of threads, that is called a \textit{threadblock} in CUDA and a \textit{workgroup} in OpenCL. We include that notion in the programming model and use the OpenCL word \textit{workgroup}. Threads in a single workgroup are run on the same processor core (by the core we understand a standard CPU core, an NVIDIA streaming multiprocessor, an OpenCL computing unit). They share access to explicitly available fast memory (that we call further \textit{shared} memory, adopting the CUDA convention, as opposed to using the OpenCL notion of \textit{local} memory) and can be explicitly synchronised using the barrier construct.

Threads from different workgroups are not synchronised and form a space of threads, executing concurrently a part of the code constituting a (computational) kernel. We assume that the whole code, in our case a code performing finite element simulations, that contains complex and large data structures, is executed in the traditional way on standard CPUs with only several kernels \textit{offloaded} to (massively) multi-core accelerators.

The act of offloading is not a necessary step in executing codes in our model. For future architectures, having standard memories accessible to massively multi-core processors with SIMD capabilities, it can be eliminated, with kernels denoting at that moment only carefully parallelized, vectorised and optimized parts of codes.

A programmer in our model can explicitly specify a storage location for selected kernel variables, as global 
or shared memory. In that case, specific, hardware dependent, considerations has to be performed to take into account, and possible optimize, different ways of accessing data in different levels of memory hierarchy. More specifically we assume that the accesses from different threads in the same SIMD group executing a single memory operation can be organized in such a way that they concern subsequent memory locations in memory. If two levels of memory are involved, the rule of subsequent accesses is considered more important for the slower memory.

The above, simple, guideline should work well for GPUs, where it should allow for forming fast \textit{coalesced} memory accesses to global memory \cite{CUDA_guide,OpenCL}. When it cannot be applied, more complex rules for performing coalesced global memory accesses and avoiding bank conflicts for shared memory should be applied. Also for CPU cores, the guideline should allow for fast vectorised memory accesses. If it turns out that the compiler and the hardware cannot perform such operations, more traditional approaches, exploiting spatial and temporal locality for each thread can be exploited.


The variables with no explicit storage location form a set of automatic variables, that are private to each thread. We assume that actual calculations are always performed using a set of registers available to a thread. If the set is large enough, all automatic variables can be stored in registers. This is the optimal situation, that can often happen for simple kernels executed on GPUs. When this is not the case, register spilling occurs, the values are stored in different levels of cache memory and eventually can be transferred, with relatively high cost, to global memory.

\subsection{Hardware platforms}
Optimal implementation of an algorithm has to consider specific features of hardware platforms for which it is designed. In the paper we try to find, for the particular problem of finite element numerical integration, the most important dependencies between the processor architectures and the performance of codes. We consider three types of computing devices, popular in HPC and having a strong presence on the current Top500 lists \cite{top500}. We do not describe the platforms in details, but compare several characteristics, important, in our view, for execution of numerical integration procedures (and definitely other scientific codes as well) that are further exploited when optimizations of these procedures are analysed. 

The first platform is a standard contemporary multi-core processor, represented by Intel Xeon E5-2620. It has 6 Sandy Bridge cores, each of which supports two-way multithreading. The second platform, an Intel Xeon Phi 5110P accelerator card has 60 cores, derived from Pentium architecture and supporting 4-way multithreading  (our OpenCL implementations can use 59 cores). The third platform is an NVIDIA Tesla K20m accelerator card with Kepler (GK110) GPU. 

\begin{table}[t]
\begin{center}
\caption{Characteristics of processors used in computational experiments.
}
\label{processors}
\begin{tabular}{|l|c|c|c|}
\hline
& \multicolumn{3}{|c|}{Processor} \\
\hline
  									 & Tesla & Xeon 	& Xeon	 \\ 
  										& K20m	& Phi	& E5-2620	 \\ 
  									 & (GK110)  & 5110P & 	($\times$2) \\ 
\hline
Number of cores	 					& 13 	& 60 (59) &	2$\times$6 \\ 
\hline
\hline
 \multicolumn{4}{|c|}{Core characteristics}\\
\hline
Number of SIMD lanes (SP/DP) 		 & 192/64 & 16/8 &	8/4		 \\ 
\hline
Number of registers 				 & 65536x32bit & 4x32x512bit & 	2x16x256bit \\ 
\hline
Shared memory (SM) size [KB] 	 & 16 or 48 & 32 (?) & 	32 (?) \\ 
\hline
L1 cache memory size [KB] 			 & 64$-$SM size & 32 & 	32 \\ 
\hline
L2 cache memory size [KB] 			 & $\approx$118 & 512 & 256	 \\ 
\hline
\end{tabular}
\end{center}
\end{table}

Table \ref{processors} compares several important characteristics of the considered processors. The data are selected to be important for the performance oriented programmers and reflects purely hardware features and some details of programming environments. By the core, as it was already mentioned, we denote a CPU core or a compute unit (streaming multiprocessor) for GPUs. For certain computing environments (as e.g. OpenCL that we use in numerical tests) the fact of supporting multithreading is reflected by multiplying the number of compute units (Intel OpenCL compiler reports the number of computing units for Xeon Phi equal to 236, and for Xeon E5-2620 equal to 12). Since hyperthreading does not necessarily lead to proportional performance improvement, we leave  in the table the number of physical cores for the Xeon processors.

The notion of SIMD lanes, in the meaning that we use in the paper, reflects the ability of a hardware to perform floating point operations (single or double precision), as it is revealed to programmers through the speed of execution of suitable low level instructions. The notion of SIMD lanes is directly related to the notion of maximal floating point performance, expressed in TFlops, that can be computed by multiplying the maximal number of operations completed by a SIMD lane in a single cycle (usually 2 for FMA or MAD operations) by the number of SIMD lanes and the frequency of core operation in THz. For CPU cores (in Xeon and Xeon Phi processors) SIMD lanes form parts of vector units, for NVIDIA GPUs SIMD lanes correspond to CUDA cores and double precision floating point units. 

The number of registers, reveals the number reported in instruction sets and programming manuals. The actual performance of the codes depends on this number but also on the number of physical resources associated (e.g. in standard CPU cores, the number of physical registers is larger than the number available in instruction sets due to the capabilities of out-of-order execution).

The picture of fast memory resources for contemporary processors is complex. Not only the details of hardware implementation (the number of banks, bus channels, communication protocols), but also the configurations change from architecture to architecture. For GPUs shared memory forms a separate hardware unit (but in the case of NVIDIA GPUs its size can be selected by the programmer). For standard CPUs, even if the programming model includes the notion of shared memory, it can be neglected in the actual execution (as is the case e.g. for Intel OpenCL compiler that, reportedly, maps it to standard DRAM memory and uses standard for CPU cache hierarchy mechanisms \cite{OpenCL_on_Xeon_Phi} -- the reason for placing ''?'' in Table \ref{processors}).

The practical computing power of processors is related to the number of SIMD lanes but
there may exist important differences between the different types of processor cores: each one can require a different number of concurrently executing threads (per single SIMD lane) to reach its maximal practical performance (this fact concerns also the performance of memory transfers). For standard CPU cores (as in the Xeon processor), one thread per single lane can suffice to reach the near peak performance. This is related to the complex architecture of cores with out-of-order execution and hardware prefetching. For simpler Xeon Phi cores, 4 threads are necessary to reach the peak performance. Even more threads are required for NVIDIA GPUs: usually several threads are sufficient to hide arithmetic operation latencies, while several tens of threads may be needed (in the case when the threads do not issue concurrent memory requests) to allow for hiding the latencies of DRAM memory operations \cite{Wong_demystifying,Owens_model}.

These facts have important consequences for designing high performance codes for the processors. The speed of execution strongly depends on the speed of providing data from different levels of memory hierarchy (registers, caches, DRAM). It is better to use resources that are faster, but their amount is limited for each core. Table \ref{memory_resources} presents the resources available to a single thread when the number of threads per SIMD lane is equal 1 for standard Sandy Bridge core in the Xeon CPU  and 4 for Xeon Phi and Kepler GPU. The number of threads per SIMD lane is selected to some extent arbitrarily, but can be understood as the minimal number of threads when high performance of execution is sought. 

It can be seen that the processors exhibit different characteristics. Standard Xeon cores (Sandy Bridge in this example) have relatively large caches with less registers. Xeon Phi cores, that are based on previous Intel CPU designs, still have large caches (although smaller than for Sandy Bridge), but, according to AVX standard, possess twice as much registers. Kepler streaming multiprocessors offer relatively large number of registers with relatively small fast memory (especially L2). If software requirements, for register per thread or shared memory per workgroup of threads, are large, then the GPU, either reduces the number of concurrently executed threads (that usually decreases the performance) or even cannot execute the code at all. 

All considered processors exhibit complex behaviour with respect to using the memory hierarchy. First generations of GPUs had not caches and, in order to get maximum performance, the programmers had to ensure that the number of requested registers did not exceed the hardware limits and that the DRAM memory accesses strictly adhered to the rules of coalescing, specific to different generations of GPU architectures. Current architectures, due to the use of caches, may not result in excessive penalties for not obeying the two mentioned above principles. This makes the picture of GPU performance less clear: sometimes it is better to allow for register spilling (when due to the too large number of requested registers the data has to be stored in cache), if this decreases the requirements for shared memory, allowing more threads to run concurrently.


\begin{table}[t]
\begin{center}
\caption{Memory resources per single thread, in terms of the number of double precision data that can be stored, for the processors used in the study (1 thread per SIMD lane for Xeon CPU and 4 threads per SIMD lane for Xeon Phi and Kepler GPU assumed -- explanation in text).}
\label{memory_resources}
\begin{tabular}{|l|c|c|c|}
\hline
& \multicolumn{3}{|c|}{Processor} \\
\hline
  									 & Tesla & Xeon 	& Xeon	 \\ 
  										& K20m	& Phi	& E5-2620	 \\ 
  									 & (Kepler GK110)  & 5110P & 	($\times$2) \\ 
\hline
Number of DP registers 				 & 128 & 32 & 	16 \\ 
\hline
Number of DP entries in SM 	 & 8 or 24 & 128 (?) & 	1024 (?) \\ 
\hline
Number of DP entries in L1 cache 			 & 32$-$SM number & 128 & 1024 \\ 
\hline
Number of DP entries in L2 cache 			 & $\approx$59 & 2048 & 8192 \\ 
\hline
\end{tabular}
\end{center}
\end{table}

\section{Performance analysis and optimization of numerical integration kernels}

\subsection{Memory requirements}

\begin{table}
\begin{center}
\caption{The number of shape  functions $N_{S}$, the  number of Gaussian integration points $N_Q$ and the associated memory requirements for the arrays appearing in numerical integration kernels for first order approximations, two popular types of finite elements: tetrahedral (tetra) and prismatic (prism) and two types of problems: with Laplace operator (Poisson) and with all convection-diffusion-reaction terms (conv-diff)}
\label{memory_requirements}
\begin{tabular}{|l|r|r|r|r|}
\hline
&  \multicolumn{4}{|c|}{Type of problem:}  \\
&   \multicolumn{2}{|c|}{Poisson} &  \multicolumn{2}{|c|}{conv-diff} \\
\hline
&  \multicolumn{4}{|c|}{Type of element:}  \\
&  \multicolumn{1}{|c|}{tetra} &  \multicolumn{1}{|c|}{prism} 
&  \multicolumn{1}{|c|}{tetra} &  \multicolumn{1}{|c|}{prism} \\
\hline
 $N_{S}$
&4&		6&	4&		6		\\
\hline
 $N_Q$
&4&	6&	4&	6\\
\hline
Integration data (reference element) & 16 & 24 & 16 & 24 \\
\hline
\hline
\multicolumn{5}{|c|}{Input/output data for each finite element}  \\
\hline
Geometry data (input) & 12 & 18 & 12 & 18 \\
\hline
PDE coefficients (sent as input) & 4 & 6 & 20 & 20 \\
\hline
Stiffness matrix ${\bfA}^{e}$ (output) & 16 & 36 & 16 & 36 \\
\hline
Load vector ${\bfb}^{e} $  (output) & 4 & 6 & 4 & 6 \\
\hline
\hline
\multicolumn{5}{|c|}{Data for single integration point -- QSS algorithm:}  \\
\hline
PDE coefficients $\bfc$ and $\bfd$ & 1 & 1 & 20 & 20 \\
\hline
Shape functions and derivatives $\bfphi$ & 16 & 24 & 16 & 24\\
\hline
Total (including ${\bfA}^{e}$ and ${\bfb}^{e} $) & \textbf{37} & \textbf{67} & \textbf{56} & \textbf{86}\\
\hline
\hline
\multicolumn{5}{|c|}{Data for all integration points -- QSS algorithm:}  \\
\hline
PDE coefficients $\bfc$  and $\bfd$ & 4 & 6 & 20 & 20 \\
\hline
Shape functions and derivatives $\bfphi$ & 28 & 144 & 28 & 144 \\
\hline
Total (including ${\bfA}^{e}$ and ${\bfb}^{e} $) & \textbf{52} & \textbf{192}& \textbf{68}& \textbf{206} \\
\hline
\hline
\multicolumn{5}{|c|}{Data for all integration points -- SSQ algorithm:}  \\
\hline
PDE coefficients $\bfc$ and $\bfd$ & 4 & 6 & 20 & 20 \\
\hline
Shape functions and derivatives $\bfphi$ & 8 & 8 & 8 & 8 \\
\hline
Total (including one entry from ${\bfA}^{e}$ and ${\bfb}^{e} $) & \textbf{14} & \textbf{16} & \textbf{30} & \textbf{30}\\
\hline
\end{tabular}
\end{center}
\end{table}

We start the performance analysis of numerical integration kernels by specifying the memory requirements that can be compared with the resources offered by different processors. The requirements determine the possible placement of data in different levels of memory hierarchy and, in consequence, the number of accesses to different levels of memory and the number of operations performed during execution.

From now on, in the rest of the paper, we concentrate on four special cases, that present important alternatives related to implementation of numerical integration. We study two types of elements: tetrahedral (geometrically linear) and prismatic (geometrically multi-linear) and two types of problems (both scalar): with Laplace operator on the left hand side and, hence, no PDE coefficients corresponding to computing stiffness matrix entries and with the full set of coefficients $\bfc$ (as can happen e.g. for different varieties of convection-diffusion-reaction equations). For the first case, denoted later on as ''Poisson'' we consider a varying right hand side term, that appears in the numerical integration as a single separate coefficient for each integration point. For the second case, denoted ''conv-diff'' the coefficients  $\bfc$ and $\bfd$ are assumed to be constant for the whole element.

Table \ref{memory_requirements} presents the basic parameters for the four considered cases of numerical integration and the sizes (in the number of scalar values) for the arrays that appear in calculations. 
The data shows a broad range of requirements from several tens of scalars to more than two hundreds (these numbers can grow to several hundreds if e.g. non-constant PDE coefficients are considered). 

The smallest requirements are associated with SSQ versions of algorithms. This can lead to the shortest execution times, however, for prismatic elements the SSQ version is accompanied by redundant calculations of Jacobian terms. On the other hand, the QSS versions, that avoid this redundancy require more resources. The SQS versions are in between, with registers required at most for a single row of stiffness matrix and with Jacobian terms calculations repeated for each row.

When comparing the numbers in Table \ref{memory_requirements} with the resources offered by the processors for a single thread, specified in Table \ref{memory_resources}, it can be seen that for some cases and some processors it could be possible to store all data necessary for numerical integration in registers. For other cases more levels of memory hierarchy has to be used. The actual placement of data will depend on the version of the algorithm and optimizations performed. 

\subsection{Input and output arrays}

Because of the importance of coalescing memory accesses to DRAM memory for graphics processors, the organization of input and output arrays can have significant influence on performance of numerical integration kernels.  

The arrays passed as arguments to the procedure (i.e. geometrical data and PDE coefficients) can be either used directly in calculations or first retrieved to shared memory or registers and than used many times in actual calculations. The first option can work well for CPU cores, for which the number of registers is relatively small, the compilers perform aggressive optimizations and the data is automatically placed in different levels of cache memory. For GPUs, the level of automation for achieving optimal performance of memory accesses is lower, and usually the programmers are responsible for ensuring the optimal organization of accesses. We adopt this point of view and explicitly design kernels with the minimal number of global memory accesses and, whenever possible, coalesced way of accessing DRAM by different threads. This means that we adopt only the second strategy, with the input data rewritten to local arrays, stored in registers or shared memory. In order to maintain the portability of kernels, we use this strategy for both CPU cores and GPUs, assuming that CPU compilers can perform the proper optimizations of kernels despite this additional step.  

The process of rewriting input data still can have several different forms. When threads rewrite subsequent data to registers, each thread refers to subsequent data concerning the element it currently processes (because of the one element--one thread parallelization strategy). In order to ensure coalesced accesses for threads in a single SIMD group, the data for different elements should be placed in subsequent memory locations in DRAM memory, while data for a single element should be spread over distant memory locations. The part of the code responsible for passing input arguments should write properly input arrays, taking into account such parameters as the size of SIMD groups, the number of threads etc. We consider this as too far-reaching requirements and assume that the data is passed in a form of arrays in a ''natural'' order, i.e. the data for a single element are placed together. 

\begin{figure}
\centering
\subfloat[]{\includegraphics[width=0.95\hsize]{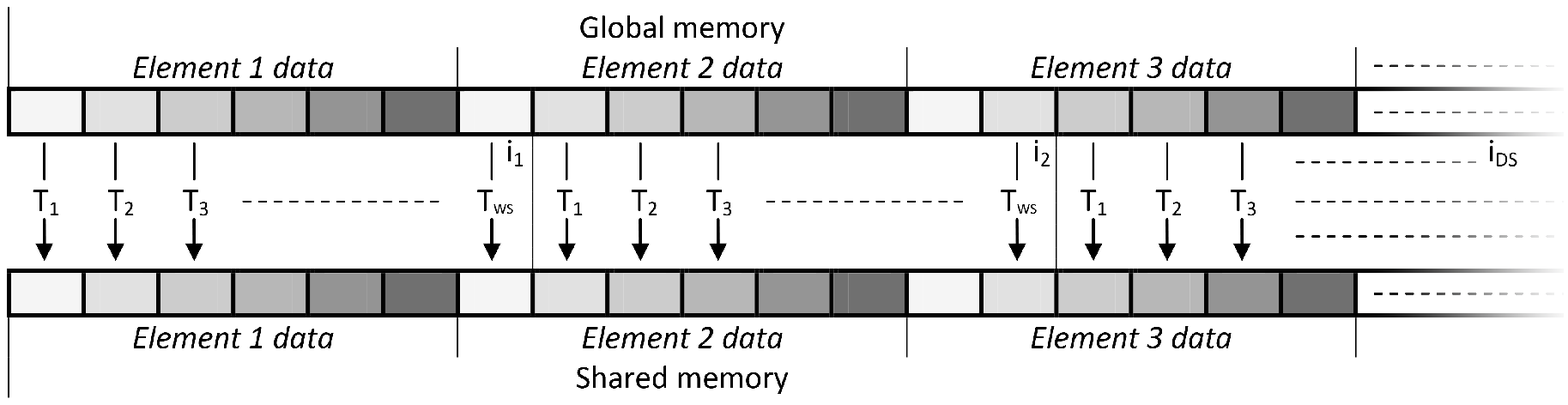}} \\
\subfloat[]{\includegraphics[width=0.95\hsize]{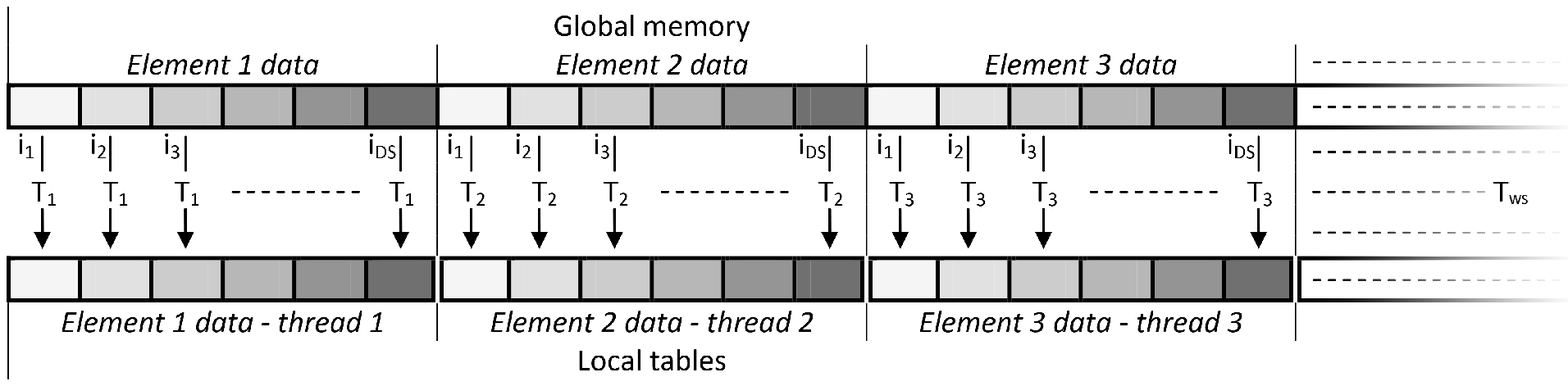}}
\caption{Organization of reading input data for numerical integration: a) coalesced reading, b) non-coalesced reading. Subsequent threads in a workgroup with size \emph{WS} are denoted by $\mathrm{T_{j}, j = 1, .., WS}$, while $\mathrm{i_k, k=1,.., DS}$ denote subsequent iterations performed by a single thread while reading a data array of size \emph{DS} }
\label{coalesced_input}
\end{figure}

In this case, when rewriting data directly to registers the accesses to global memory are not coalesced. The typical remedy to that, used often for GPUs, is to employ shared memory as a temporary storage with shared access for threads from a single workgroup. The threads read input data in a coalesced way, with subsequent threads from a single workgroup referring to subsequent memory locations, and than write to shared memory, as is illustrated in Fig.~\ref{coalesced_input}. The order of writing to and reading from shared memory is assumed not to have the influence on performance and is not optimized (the order of data is left the same as in DRAM memory).  

The data placed in shared memory can be left there and later used in calculations, or can be rewritten to registers, with shared memory possibly reused for other data. Both options are taken into account when considering the optimization of kernels. 

\begin{figure}
\centering
\subfloat[]{\includegraphics[width=0.95\hsize]{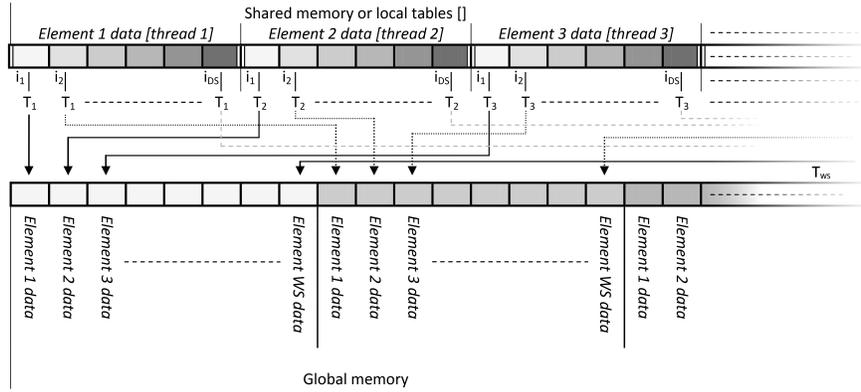}} \\
\subfloat[]{\includegraphics[width=0.95\hsize]{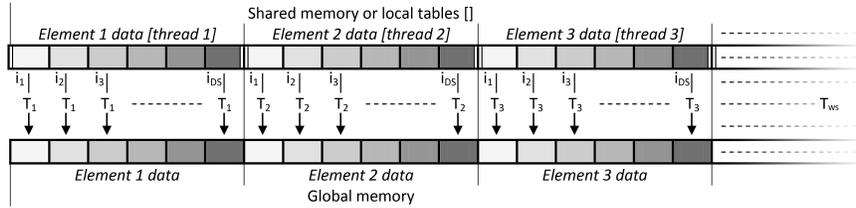}}
\caption{Organization of writing output data for numerical integration: a) coalesced writing, b) non-coalesced writing.  Subsequent threads in a workgroup with size \emph{WS} are denoted by $\mathrm{T_{j}, j = 1, .., WS}$, while $\mathrm{i_k, k=1,.., DS}$ denote subsequent iterations performed by a single thread while writing a data array of size \emph{DS}}
\label{coalesced_output}
\end{figure}

We adopt a different strategy for output data. Since we leave produced stiffness matrices and load vectors for further use by other kernels or standard CPU code, we assume that the procedures that read them can adapt to the format determined by integration kernels. We also design the versions with coalesced and non-coalesced memory accesses, but now the versions with coalesced memory accesses do not use additional shared memory buffers and write the data directly from data structures related to threads (that can be stored either in registers or in shared memory). As a consequence the data for single element are now spread over distant memory locations and the procedures that read them must take into account the particular formats, that depend on details of the organization of calculations. For non-coalesced writing, a single thread writes the data related to a single element to subsequent memory locations -- the situation that can be advantageous for CPU cores with not vectorized memory operations.
The situations of coalesced and non-coalesced memory accesses of output data are illustrated in Fig.~\ref{coalesced_output}.

\subsection{Shared memory usage}

The algorithms presented so far can be used to create a proper code for compilers in different software development kits for multi-core processors. For architectures based on standard CPU cores, the compilers can produce codes that can reach high performance, thanks to different automatic optimizations. For GPUs, however, there exists one more important optimization, that can have substantial impact on performance and that has to be taken into account explicitly in the source code. 

This optimization is the use of shared memory, either as a mean for inter-thread communication or an explicitly manageable cache. The first option has been already investigated in the previous section, for the purpose of optimal reading of input data. The second option is related to the register requirements of the kernels. If this requirements are too large, the variables designed as local (private to each thread) are spilled to caches and DRAM memory. The last situation, that can happen often due to small cache resources of GPUs, can lead to severe performance penalties. In order to relieve register pressure some of data used in calculations can be placed explicitly in shared memory. Shared memory accesses are usually several times slower than register accesses but, in turn, several times faster then (even coalesced) DRAM accesses. 

One thing has to be taken into account for such designs. When one variable is designed to be placed in shared memory instead of a register, the kernel has to allocate the place in shared memory for all threads from a single workgroup. Storing an array with $k$-entries for a single thread in shared memory, requires the allocation of $k$ times the size of workgroup entries. Since workgroups sizes for GPUs are designed as multiples of SIMD groups sizes (with the minimum recommended both for AMD and NVIDIA GPUs equal to 64), moving data to shared memory involves relatively large resource requirements. Because of that, when designing numerical integration kernels, we consider the placement in shared memory of only one array appearing in final calculations of stiffness matrix and load vector entries, either the one for PDE coefficients or for shape functions and their derivatives or, eventually, for the output ${\bfA}^{e}$ and ${\bfb}^{e} $ matrices. 

It is difficult to investigate all possibilities for placement of data in shared memory (including the options of reusing shared memory after coalesced reading for some other (than input) arrays). The number of options can reach tens (or even hundreds if the assumption of placing only one array in shared memory is abandoned). The factors that influence the performance in these cases include, apart from the number of accesses to shared memory, that can be induced e.g. from the analysis of source code, also the relative sizes and speeds of different levels of memory, as well as compiler strategies for allocating registers, that influence the number of accesses to caches and global DRAM memory. These factors can differ much for different architectures. 

Because of that, in order to select the best option, we choose a popular in HPC strategy of parameter based tuning. We create a single code with options concerning the placement of particular arrays in shared memory, that can be switched on or off at compile time, and perform an exhaustive search of the whole space of options by running several tens of variants of the procedure. 

\subsection{Operation count}

The operations performed during kernel executions include global memory accesses, shared memory accesses and the remaining operations. Among the latter we are interested only in floating point operations. We base our execution time estimates, developed always for a single finite element, on the numbers of global memory accesses and floating point operations, due to the difficulties with taking into account shared memory and cache accesses.

When performing analysis at this level, the final optimizations, performed by software (compilers) and hardware should be taken into account. This can be difficult, especially when different architectures are taken into account. We try to develop the estimates, considering such classical optimizations as loop invariant code motion, common subexpression elimination, loop unrolling, induction variable simplification, etc. (that as we mentioned earlier are not directly introduced into the source code, but left for the compilers). 
In the estimates we accept the minimal number of performed operations and, when calculating performance data for processors, we base them on the estimated numbers.

The number of global memory accesses is relatively easy to estimate.
As was already mentioned, in our implementations we accept the principle of explicitly managing the accesses to global memory, i.e. we do not use global arrays in calculations but rewrite the data to shared memory or registers (local, automatic variables) and than use the latter in computations. Hence, the number of global memory accesses can be estimated based on the sizes of input and output arrays for the numerical integration procedures (Table \ref{memory_requirements}). All variants (QSS, SQS and SSQ) perform the same number of global memory accesses reported in the first line of Table \ref{accesses_and_operations}.

\begin{table}
\begin{center}
\caption{The number of global and shared memory accesses (for shared memory there are considered only additional accesses not related to reading data from or writing data to global memory), the number of operations and the resulting arithmetic intensities for numerical integration kernels for first order approximations, two popular types of finite elements: tetrahedral (tetra) and prismatic (prism) and two types of problems: with Laplace operator (Poisson) and with all convection-diffusion-reaction terms (conv-diff)}
\label{accesses_and_operations}
\begin{tabular}{|l|r|r|r|r|}
\hline
&  \multicolumn{4}{|c|}{Type of problem:}  \\
&   \multicolumn{2}{|c|}{Poisson} &  \multicolumn{2}{|c|}{conv-diff} \\
\hline
&  \multicolumn{4}{|c|}{Type of element:}  \\
&  \multicolumn{1}{|c|}{tetra} &  \multicolumn{1}{|c|}{prism} 
&  \multicolumn{1}{|c|}{tetra} &  \multicolumn{1}{|c|}{prism} \\
\hline
\hline
\multicolumn{5}{|c|}{The number of global memory accesses:}  \\
\hline
all variants
&36&		66&	52&		80		\\
\hline
\hline
\multicolumn{5}{|c|}{The number of auxiliary shared memory accesses -- QSS algorithm}  \\
\hline
Geometry data in shared memory & 12 & 108 & 12 & 108 \\
\hline
PDE coefficients $\bfc$ in shared memory & 4 & 6 & 80 & 120 \\
\hline
Shape functions $\bfphi$ in shared memory & 60 & 210 & 60 & 210 \\
\hline
${\bfA}^{e}$ and ${\bfb}^{e} $ in shared memory & 180 & 546 & 180 & 546 \\
\hline
\hline
\multicolumn{5}{|c|}{The number of operations}  \\
\hline
QSS algorithm & 290 & 2700 & 986 & 4806 \\
\hline
SQS algorithm & 290 & 10416 & 986 & 12492 \\
\hline
SSQ algorithm & 290 & 54876 & 1623 & 65232 \\
\hline
\hline
\multicolumn{5}{|c|}{Arithmetic intensities}  \\
\hline
QSS algorithm & 8 & 41 & 19 & 60 \\
\hline
SQS algorithm & 8 & 158 & 19 & 156 \\
\hline
SSQ algorithm & 8 & 831 & 31 & 815 \\
\hline
\end{tabular}
\end{center}
\end{table}

\begin{table}[t]
\begin{center}
\caption{Performance characteristics of processors used in computational experiments.
}
\label{processors_performance}
\begin{tabular}{|l|c|c|c|}
\hline
& \multicolumn{3}{|c|}{Processor} \\
\hline
  									 & Tesla & Xeon 	& Xeon	 \\ 
  										& K20m	& Phi	& E5-2620	 \\ 
  									 & (GK110)  & 5110P & 	($\times$2) \\ 
\hline
Peak DP performance [TFlops] 		 & 1.17 & 1.01 (0.99) & 2$\times$0.096	 \\ 
\hline
Peak memory bandwidth [GB/s] 		 & 208 & 320 & 	2$\times$42.6 \\ 
\hline
Limiting arithmetic intensity (DP) 	& 45 & 25 & 18 \\
\hline
\hline
Benchmark (DGEMM) performance 		 & 1.10 & 0.84 & 0.18	 \\ 
\hline
Benchmark (STREAM) bandwidth 		 & 144 & 171 & 	67 \\ 
\hline
Limiting arithmetic intensity (DP) & 61 & 39 & 21 \\
\hline
\end{tabular}
\end{center}
\end{table}

The situation is much more complex for shared memory accesses and floating point operations. First, for different variants of kernels there will be different organization of operations.  Moreover, for different organizations compilers can perform different sets of optimizations. We present the estimated numbers for shared memory accesses and floating point operations in Table \ref{accesses_and_operations}. In deriving the estimates we tried to take into account the possible optimizations, related e.g. to  the facts that certain values are constant during calculations (e.g. derivatives of shape functions for different integration points within the tetrahedral element, PDE coefficients for different integration points in the convection-diffusion case) and that the resulting stiffness matrices may be symmetric.

Calculations in all versions of numerical integration algorithm contain several phases, such as:
\begin{itemize}
\item
computing real derivatives of geometrical shape functions (9 operations for tetrahedral elements, 126 operations per integration point for prisms)
\item
computing Jacobian terms (49 operations in all cases, but performed once for tetrahedra and repeated for each integration point for prisms)
\item
computing real derivatives for shape functions (60 operations for tetrahedra and 90 operations per integration point for prismatic elements)
\item
calculating the entries for the stiffness matrix and the load vector
\end{itemize}

The final calculations are performed in a triple loop over integration points and shape functions, but in many cases the loops are fully unrolled by the compiler and the resulting numbers of operations are the same for all variants of numerical integration. This happens for tetrahedral elements due to many calculations moved outside all loops (since derivatives of shape functions are constant over the element). For prismatic elements the fact that several calculations must be repeated for each integration point results in much larger number of operations for the SQS and especially SSQ versions of the algorithm, as compared to the QSS version. 

Based on the estimated numbers of global memory accesses and floating point operations we calculate arithmetic intensities (defined as the number of operations performed per global memory access) for the different variants of integration kernels and different cases considered. The calculated values, reported in Table \ref{accesses_and_operations}, can be compared with performance characteristics of the processors used in the study. We present such characteristics in Table \ref{processors_performance}.
For each processor we report its peak performance and benchmark performance for the two types of operations -- memory accesses and floating point operations (as benchmarks we consider STREAM for  the memory bandwidth and dense matrix-matrix multiplication (DGEMM) for  floating point performance; moreover from now on we restrict our analysis to only double precision calculations, as more versatile than single precision). Based on these data and assuming the simple roofline model of processor performance \cite{Williams_roofline_2009}, the limiting arithmetic intensity, i.e. the value of arithmetic intensity that separates the cases of the maximal performance limited by memory bandwidth from that limited  by the speed of performing arithmetic operations (pipeline throughput), is calculated for each processor and theoretical and benchmark performance.

Comparing the data in Table \ref{accesses_and_operations} and Table \ref{processors_performance} we see that the performance when running different variants of numerical integration kernels can be either memory bandwidth or processor performance (pipeline throughput) limited. For tetrahedral elements the performance for both example problems is memory limited for all architectures (excluding the not optimal SSQ version). For prismatic elements, even when benchmark (i.e. higher) limiting arithmetic intensity is considered, the performance of kernel execution should be pipeline limited for Xeon and Xeon Phi architectures. For the Kepler architecture and the optimal QSS variant the performance will be memory limited for Poisson problem, while is on the border between being memory or pipeline limited for the convection-diffusion example. For other variants the performance is pipeline limited but the large number of operations in these cases makes them far from optimal.

\subsection{Execution time}
Based on the data in Tables \ref{accesses_and_operations} and \ref{processors_performance} we perform final estimates of execution time for different variants of numerical integration and different architectures. We calculate separately the estimated time determined by the number of global memory accesses and the benchmark (STREAM) memory bandwidth of processors and the time determined by the number of floating point operations and the arithmetic throughput in DGEMM. Table \ref{execution_times} contains the longer times from the two, computed as the estimate for each architecture.

The values in Table  \ref{execution_times} form lower bounds for the times possible to achieve. There are many factors that can slow down the execution of algorithms. The most important include: 
\begin{itemize}
\item
too low occupancy for GPUs (too small number of concurrently executing threads in order to effectively hide instruction or memory latencies) -- caused by excessive requirements for registers or shared memory
\item
shared and cache memory accesses for GPUs, not considered in Table  \ref{execution_times} and caused by either explicit use of shared memory or register spilling to caches (or even to global memory)
\item
using other than FMA or MAD instructions -- peak performance is obtained by all the considered processors for algorithms using almost exclusively FMA or MAD operations that double the performance as compared to using other floating point instructions
\item
employing barriers to ensure the proper usage of shared memory
\end{itemize}

\begin{table}
\begin{center}
\caption{Estimates for the execution time of numerical integration kernels for first order approximations, two popular types of finite elements: tetrahedral (tetra) and prismatic (prism), two types of problems: with Laplace operator (Poisson) and with all convection-diffusion-reaction terms (conv-diff) and three example processors representing Xeon, Xeon Phi and Kepler architectures (times are reported in nanoseconds for a single element)}
\label{execution_times}
\begin{tabular}{|l|r|r|r|r|}
\hline
&  \multicolumn{4}{|c|}{Type of problem:}  \\
&   \multicolumn{2}{|c|}{Poisson} &  \multicolumn{2}{|c|}{conv-diff} \\
\hline
&  \multicolumn{4}{|c|}{Type of element:}  \\
&  \multicolumn{1}{|c|}{tetra} &  \multicolumn{1}{|c|}{prism} 
&  \multicolumn{1}{|c|}{tetra} &  \multicolumn{1}{|c|}{prism} \\
\hline
\hline
\multicolumn{5}{|c|}{Tesla K20m (GK110 -- Kepler)}  \\
\hline
QSS algorithm & 2.00 & 3.67 & 2.89 & 4.44 \\
\hline
SQS algorithm & 2.00 & 9.47 & 2.89 & 11.36 \\
\hline
SSQ algorithm & 2.00 & 49.89 & 2.89 & 59.30 \\
\hline
\hline
\multicolumn{5}{|c|}{Xeon Phi 5110P}  \\
\hline
QSS algorithm & 1.68 & 3.21 & 2.43 & 5.72 \\
\hline
SQS algorithm & 1.68 & 12.40 & 2.43 & 14.87 \\
\hline
SSQ algorithm & 1.68 & 57.87 & 2.43 & 69.40 \\
\hline
\hline
\multicolumn{5}{|c|}{Xeon E5-2620 (two socket configuration)}  \\
\hline
QSS algorithm & 4.30 & 15.00 & 6.21 & 26.70 \\
\hline
SQS algorithm & 4.30 & 57.87 & 6.21 & 69.40 \\
\hline
SSQ algorithm & 4.30 & 304.87 & 9.02 & 362.40 \\
\hline
\end{tabular}
\end{center}
\end{table}

\subsection{Transfer from and to host memory}

The last issue that we deal with, is the one that is often considered first -- the time required for transfer of data between host memory and accelerator memory (when these two memories are separate, i.e. when accelerator is connected with the CPU using PCIe bus, as is the case for Tesla K20m and Xeon Phi). 

The reason why we consider it last is that, based on the discussions performed so far, it is evident that if numerical integration is considered alone, without considering other procedures of finite element processing, the time required for transferring input and output data (or even input data alone) strongly dominates the time for performing actual computations. This becomes obvious after realizing that the performance of numerical integration kernels is usually memory bound or close to being memory bound, that the amount of data in global memory accessed by accelerator during calculations is the same as that of host-accelerator memory transfer and that the bandwidth of PCIe is several times lower than that of either GPU or Xeon Phi global memory. 

Because of that we do not discuss this issue here. Apart from the fact that such a discussion should include broader context of finite element simulations, there is another reason for not going into details. The solution with PCIe CPU-accelerator connection is universally considered as one of the most important performance bottlenecks for accelerator computing. Hardware providers are considering many different options that should change this situation. When these solutions become available, the detailed analyses of the influence of host memory-accelerator memory transfer on performance of codes can be performed (for some initial experience see e.g. \cite{cmms_APU_2015}).

\section{Computational experiments}

We test the three versions (QSS, SQS, SSQ) of numerical integration kernels  for the four described example cases (Poisson and convection-difffusion for tetrahedral and prismatic elements) and the three selected processor architectures Intel Xeon E5-2620, Intel Xeon Phi 5110P and NVIDIA Tesla K20m with Kepler GK110 architecture.
For all processors we use 64-bit Linux with 2.6.32 kernel. For Xeon processors we use Intel SDK for OpenCL Applications version 4.5 (with OpenCL 1.2 support). For NVIDIA GPU we use CUDA SDK version 5.5 with OpenCL 1.1 support and 331.20 driver. We use the same kernels for all three architectures, changing only the size of workgroups: 8 for Xeon, 16 for Xeon Phi and 64 for Kepler.

\begin{figure}
\centering
\includegraphics[width=0.9\hsize]{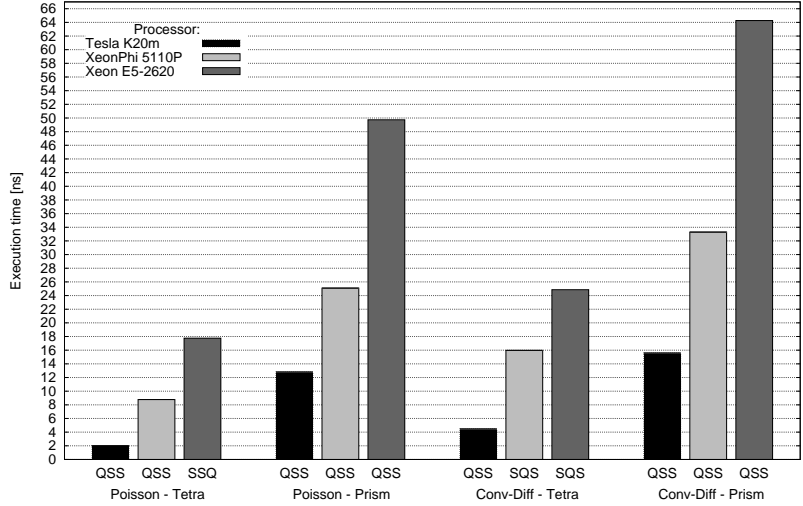}
\caption{Perfromance/Execution times of numerical integration for a single tetrahedral and prismatic element, Poisson and convection-diffusion problem and the best version among integration kernels for each of tested architectures: Xeon, Xeon Phi and Kepler (Tesla K20m)}
\label{figure_time}
\end{figure}

\begin{figure}
\centering
\subfloat[]{\includegraphics[width=0.5\hsize]{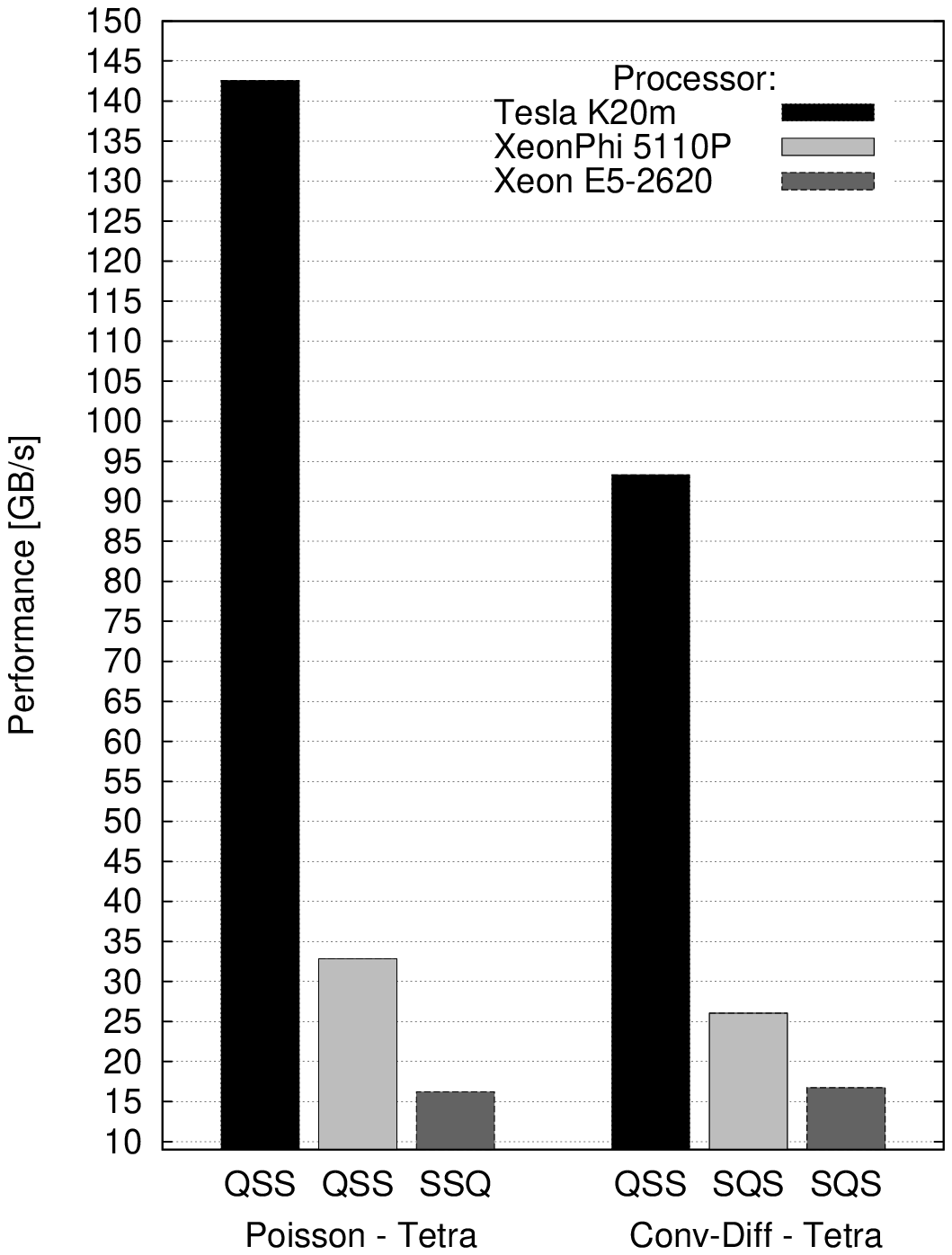}}
\subfloat[]{\includegraphics[width=0.5\hsize]{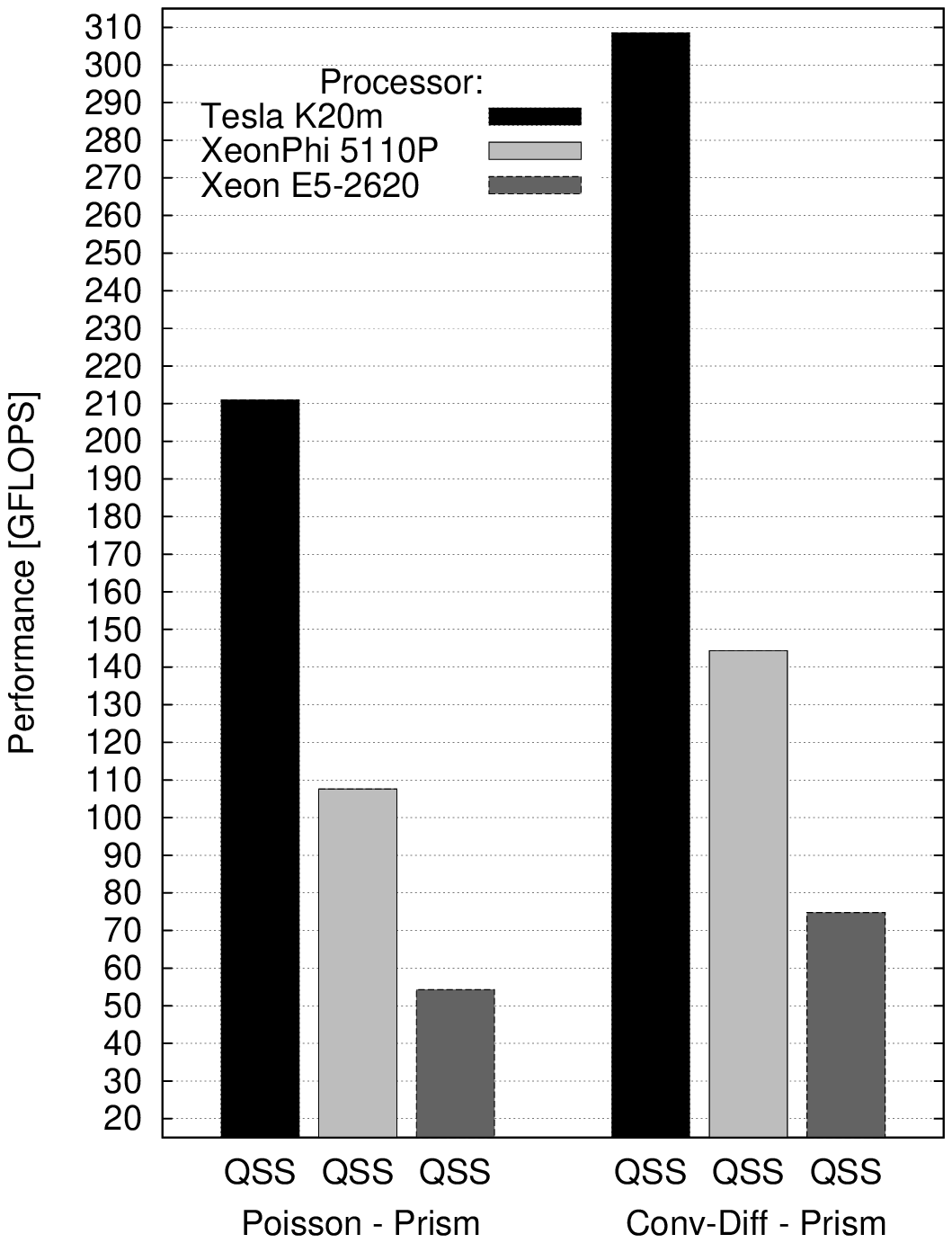}}
\caption{Absolute perfromance of numerical integration for Poisson and convection-diffusion problem and the best version among integration kernels for each of tested architectures: Xeon, Xeon Phi and Kepler (Tesla K20m): a) in GB/s for a single tetrahedral element, b) in GFLOPS for a single prismatic element, }
\label{figure_abs_perf}
\end{figure}

\begin{figure}
\centering
\includegraphics[width=0.9\hsize]{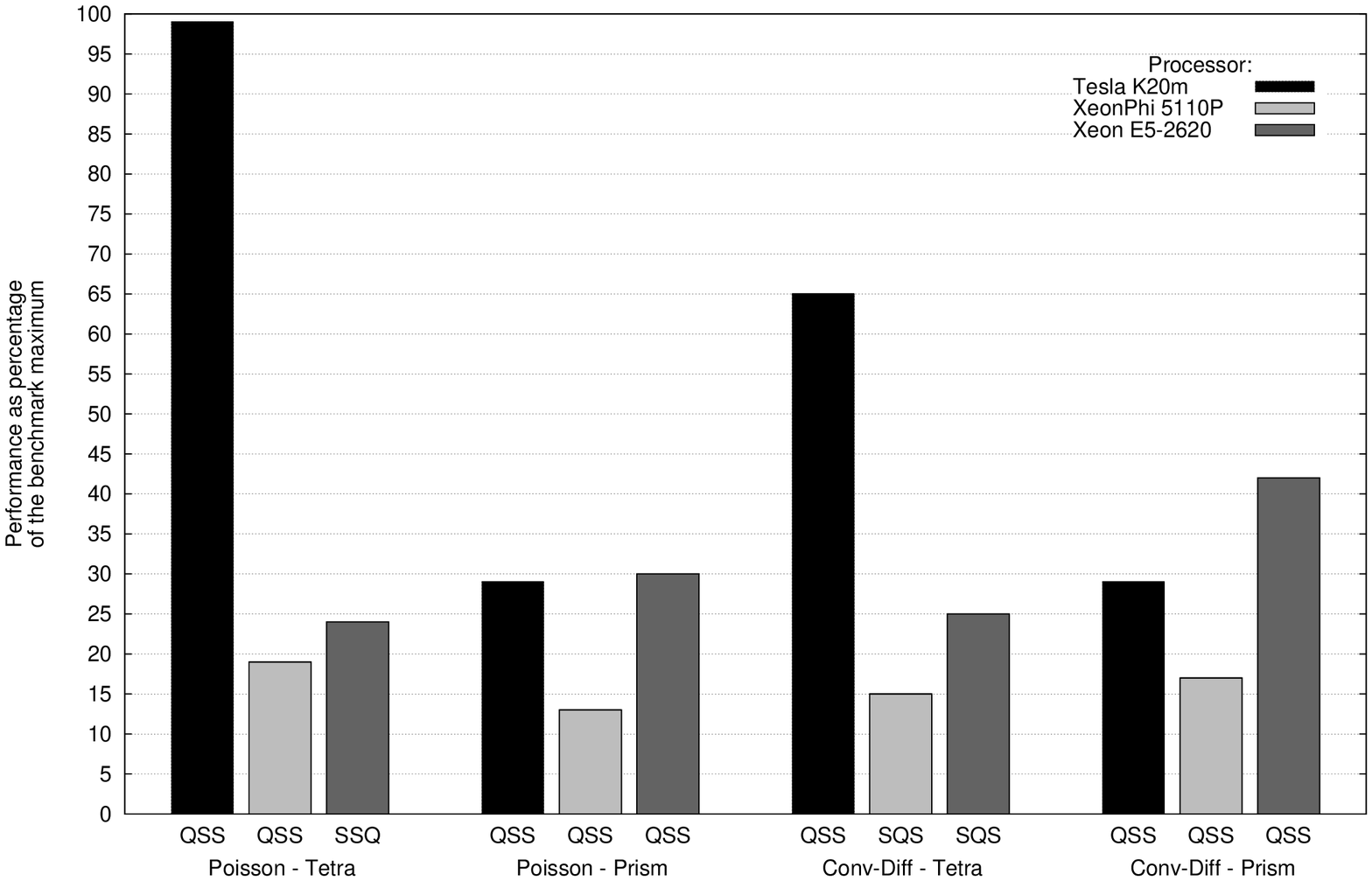}
\caption{Relative performance (as percentage of the maximal benchmark performance) of numerical integration for a single tetrahedral and prismatic element, Poisson and convection-diffusion problem and the best version among integration kernels for each of tested architectures: Xeon, Xeon Phi and Kepler (Tesla K20m). For tetrahedral elements the percentage refers to the STREAM memory bandwidth benchmark, for prismatic elements  it refers to the DGEMM floating point performance (except for Kepler architecture where it is still related to the STREAM benchmark). }
\label{figure_rel_perf}
\end{figure}

For each particular kernel and problem we perform a series of experiments with different combinations of parameters determining the use of shared memory in computations. In Table \ref{results_time} we present the best times for each kernel, indicating additionally what percentage of the best performance (associated with the times presented in Table  \ref{execution_times}) was achieved. Figures \ref{figure_time}, \ref{figure_abs_perf} and \ref{figure_rel_perf} illustrates the comparison of performance for each case of problem and approximation and different processors.

\begin{table}
\begin{center}
\caption{Experimental execution times and the achieved percentage of the best performance associated with times in Table  \ref{execution_times} for numerical integration kernels for first order approximations, two popular types of finite elements: tetrahedral (tetra) and prismatic (prism), two types of problems: with Laplace operator (Poisson) and with all convection-diffusion-reaction terms (conv-diff) and three example processors representing Xeon, Xeon Phi and Kepler architectures (times are reported in nanoseconds for a single element).}
\label{results_time}
\begin{tabular}{|l|r|r|r|r|}
\hline
&  \multicolumn{4}{|c|}{Type of problem:}  \\
&   \multicolumn{2}{|c|}{Poisson} &  \multicolumn{2}{|c|}{conv-diff} \\
\hline
&  \multicolumn{4}{|c|}{Type of element:}  \\
&  \multicolumn{1}{|c|}{tetra} &  \multicolumn{1}{|c|}{prism} 
&  \multicolumn{1}{|c|}{tetra} &  \multicolumn{1}{|c|}{prism} \\
\hline
\hline
\multicolumn{5}{|c|}{Tesla K20m (GK110 -- Kepler)}  \\
\hline
QSS algorithm & 2.02 (99\%) & 12.80 (29\%)  & 4.46 (65\%) & 15.58 (29\%) \\
\hline
SQS algorithm & 2.02 (99\%) & 38.39 (25\%) & 6.05 (48\%) & 64.81 (18\%) \\
\hline
SSQ algorithm & 2.02 (99\%) & 94.99 (53\%) & 6.13 (47\%) & 194.93 (30\%) \\
\hline
\hline
\multicolumn{5}{|c|}{Xeon Phi 5110P}  \\
\hline
QSS algorithm & 8.77 (19\%) & 25.09 (13\%) & 17.11 (14\%) & 33.29 (17\%) \\
\hline
SQS algorithm & 11.35 (15\%) & 41.20 (30\%) & 15.97 (15\%) & 60.18 (25\%) \\
\hline
SSQ algorithm & 11.43 (15\%) & 101.20 (65\%) & 16.53 (15\%) & 132.46 (59\%) \\
\hline
\hline
\multicolumn{5}{|c|}{Xeon E5-2620 (two socket configuration)}  \\
\hline
QSS algorithm & 19.31 (22\%) & 49.74 (30\%) & 26.74 (23\%) & 64.27 (42\%) \\
\hline
SQS algorithm & 18.77 (23\%) & 128.07 (45\%) & 24.86 (25\%) & 133.14 (52\%) \\
\hline
SSQ algorithm & 17.76 (24\%) & 385.33 (79\%) & 26.11 (35\%) & 475.41 (76\%) \\
\hline
\end{tabular}
\end{center}
\end{table}

The detailed results of experiments lead to several remarks belonging to different domains of interest. For practitioners performing finite element simulations on modern hardware the interesting facts include:
\begin{itemize}
\item
the execution time of finite element integration depends strongly not only on the order of approximation (this fact is obvious from the classical complexity analysis), but also on the type of element and the kind of problem solved. The differences are not so significant as in the case of different approximation orders \cite{Kirby_FEM_operators,camwa_13_ni_CBE,camwa_14_ni_GPU}, but still can reach an order of magnitude
\item
due to large differences in execution time the problem of proper mapping to hardware and optimization of numerical integration may be more or less important for the time of the whole finite element simulation, depending also e.g. on the time required by the solver of linear equations (if present) \cite{ppam07,maciek_camwa_2014_cost})
\end{itemize}

For finite element software developers, especially in the HPC domain, there are the following general observations:
\begin{itemize}
\item
large resource requirements for some variants of numerical integration (e.g. convection-diffusion problems and prismatic elements), limits the performance achieved by the GPU architecture in our comparison; the reason lies mainly in too small number of threads (SIMD groups) that can run concurrently on a single core (streaming multiprocessor)
\item
the QSS variants of the numerical integration algorithm turned out to be the most versatile and efficient; due to the proper optimizations in the source code and performed by the compilers, the performance of QSS kernels was either the fastest or slower by only several percent than the fastest version
\item
detailed results of experiments showed that the performance on our GPU (as is also the case for other contemporary GPU architectures) is sensitive to the way the data are read and written to memory, coalescing memory accesses turned out to be more important for the Kepler processor than for both Xeons
\item 
when the performance of algorithm is memory bound and there is a significant influence of the way the data are accessed in global memory (as is the case e.g. for Kepler architecture, tetrahedral element and Poisson problem), there is a growing importance of data layout in memory (and the use of output data by other components of finite element codes) \cite{Cecka_2011,Cai_matrix_free_FEM_GPU_2013,Kirby_FEM_pipeline_GPU} 
\item 
the GPU architecture turned out to be also more sensitive to the use of OpenCL shared memory (as suggested by the Intel OpenCL compiler manual \cite{opencl_phi_guide}, this had no significant impact on the performance of Xeons)
\item
together, the changes in the way data are read and written to the global memory and the changes in shared memory usage caused the differences in execution time of the order 2-3 for Xeons, and more than 5 for Kepler
\end{itemize}

In general, for some test cases, the performance of obtained portable numerical integration kernels can be considered satisfactory. Still, for each architecture and most of the cases considered, the benchmark results suggest that it should be possible to achieve execution times several times shorter (from 3 times for Kepler architecture, up to almost 8 times for Xeon Phi). This could however usually be obtained by either considering the specialized domain specific optimizing compilers or by loosing the portable character of the kernels and introducing architecture specific optimizations into the source code. Another possibility is to consider different parallelization strategies, the option that we plan to follow in subsequent publications.

Finally there are some observations concerning the hardware itself (these observations are common to many algorithms, but are confirmed for our example case of finite element numerical integration):
\begin{itemize}
\item
our example GPU, Tesla K20m, turned out to be the fastest processor in the comparison; the observations made by several authors \cite{Phi_pain_2013,assembly_Phi_Hanzlikova_2013,SpMV_Phi_Saule_2013}
show that Xeon Phi processors applied to different scientific computing algorithms, although usually faster than standard Xeon CPUs, are usually slower than the recent HPC targeted GPUs
\item
the detailed results of experiments for particular optimization variants of kernels showed significant resemblance of Xeon and Xeon Phi performance, indicating the similarities between these architectures.
\end{itemize}

\section{Related work}

The subject of numerical integration on modern multi-core processors, especially graphics processors, has been investigated in several contexts. The first works concerned GPU implementations for specific types of problems, such as higher order FEM approximations in earthquake modelling and wave propagation problems \cite{komatitsch09}, GPU implementations of some variants of discontinuous Galerkin approximation \cite{Klockner_2009} or higher order approximations for electromagnetics problems \cite{Dziekonski_generation}.

The second important context for which finite element numerical integration was considered, is the creation of domain specific languages and compilers. Interesting works for this approach include \cite{Markall_2010,Markall_2013,Knepley_2013,Logg_DOLFIN,Luporini_2015}. The research in the current paper can be understood as laying theoretical and experimental grounds for possible future finite element compilers, able to create optimized code for all types of problems, elements and approximations, as well as different processor architectures.

Finally, numerical integration, in more or less details, have been discussed in the context of the whole simulation procedure by the finite element codes. Some works in this area include \cite{Huthwaite_elastodynamics_GPU_2014,Idelsohn_explicit_conv_diff_GPU_2012,Goeddeke_FEM_toolkit_GPU_2013,Mafi_nonlinear_FEM_on_GPU_2014,Georgescu_FEM_on_GPU_2013,Kirby_FEM_pipeline_GPU,fem_on_gpu_reguly_giles}.
Usually more attention to numerical integration has been given in articles that consider the process of creation of systems of linear equations for finite element approximations, with the important examples such as \cite{Filipovic_Generation_nonlinear_mechanics_GPU_2012,FEM_assembly_energy_2013,Cecka_gems,Dziekonski_assembly}.

The present paper forms a continuation of our earlier works, devoted solely to the subject of finite element numerical integration. The first papers considered higher order approximations, starting with first theoretical and experimental investigations in \cite{ppam_09_gpu,ppam_09_cell} and culminating in larger articles devoted to the implementation and performance of numerical integration kernels for multi-core processors with vector capabilities \cite{camwa_13_ni_CBE} (especially IBM Power XCell processor) and GPUs \cite{camwa_14_ni_GPU}.

The investigations in the current paper on one hand touch the narrower subject of only first order approximations, but on the other hand concern two important types of finite elements (geometrically linear and non-linear) and are performed, in a unifying way, for three different processor architectures used in scientific computing. This scope, combined with the depth of investigations concerning the performance of OpenCL kernels, differentiate the article from the other on this subject.

\section{Conclusions}
We have investigated the optimization and performance on current multi- and many-core processors for an important computational kernel in scientific computing, the finite element numerical integration algorithm for first order approximation. We have used a unifying approach of OpenCL for  programming model and implementation on three popular architectures in scientific computing: Intel Xeon (Sandy Bridge), Intel Xeon Phi and NVIDIA Kepler. We have demonstrated that this approach allows for exploiting multi-core and vector capabilities of processors and for obtaining satisfactory levels of performance, as compared to the theoretical and benchmark maxima, not loosing the portability of the developed code (we have used the same OpenCL kernels for all architectures). The investigations in the paper reveal that, even for the simple problems, elements and approximations as considered in the paper, there is a significant variation of required resources and associated complexities for the algorithm, that can lead to different problems when mapping to architectures of modern processors. Nevertheless, the results of computational experiments show that for all the cases considered in the paper, the numerical integration algorithm can be successfully ported to massively multi-core architectures, and hence, when used in finite element codes should not form a performance bottleneck. 
The presented detailed analyses indicate what conditions must be met in order to obtain the best performance of the kernels and what performance can be expected when numerical integration is used on different processors for different types of problems and finite element approximations. 









\end{document}